\documentclass[conference]{IEEEtran}

\usepackage{graphicx,cite,epsfig}
\usepackage{amsmath} 
\usepackage{amssymb}  
\DeclareMathOperator *{\argmin}{argmin}
\usepackage{pifont}
\usepackage{stmaryrd}
\usepackage{bbding}
\usepackage{subfigure}
\usepackage{algorithm}
\usepackage{algorithmic}
\usepackage{array}
\usepackage{amsthm}
\usepackage{url}
\usepackage{mathrsfs}
\usepackage{color, soul} 
\usepackage{flushend}
\usepackage{mathrsfs}
\usepackage{epstopdf}

\begin{document}
\title{Learning to Flip Successive Cancellation Decoding of Polar Codes with LSTM Networks}
\author{\IEEEauthorblockN{Xianbin Wang, Huazi Zhang, Rong Li, Lingchen Huang, Shengchen Dai, Yourui Huangfu, Jun Wang}
\IEEEauthorblockA{Huawei Technologies Co., Ltd., Hangzhou, China.\\
Emails: \{wangxianbin1,zhanghuazi,lirongone.li,huanglingchen,daishengchen,huangfuyourui,justin.wangjun\}@huawei.com}}

\maketitle
\thispagestyle{empty}
\begin{abstract}
The key to successive cancellation (SC) flip decoding of polar codes is to accurately identify the first error bit.
The optimal flipping strategy is considered difficult due to lack of an analytical solution.
Alternatively, we propose a deep learning aided SC flip algorithm.
Specifically, before each SC decoding attempt, a long short-term memory (LSTM) network is exploited to either (i) locate the first error bit, or (ii) undo a previous ``wrong'' flip.
In each SC attempt, the sequence of log likelihood ratios (LLRs) derived in the previous SC attempt is exploited to decide which action to take.
Accordingly, a two-stage training method of the LSTM network is proposed, i.e., learn to locate first error bits in the first stage, and then to undo ``wrong'' flips in the second stage.
Simulation results show that the proposed approach identifies error bits more accurately and achieves better performance than the state-of-the-art SC flip algorithms.
\end{abstract}

\section{Introduction}\label{1}
Polar codes are the first capacity-achieving channel codes with low-complexity successive cancellation (SC) decoding \cite{ArikanPolar,5G}.
For polar codes of moderate length, however, the SC decoding performance is not satisfactory.
Once a wrong bit decision (bit error) occurs, it has no chance to be corrected due to the sequential bit decisions.
Successive cancellation list (SC-List) \cite{SCL} and CRC-aided SC list (CA-SC-List) \cite{CASCL} decoders address this issue.
For each bit, both 0/1 decisions are evaluated and developed as two decoding paths.
Among all paths, the $L$ most likely ones are preserved, such that the correct path survives at a higher chance.
These algorithms obtain significant performance gain at the cost of roughly $L$-times memory and complexity.

An alternative low-complexity solution is SC flip decoding \cite{SCFlip}.
Instead of keeping multiple decoding paths, only one path is developed at a time.
If an SC decoding attempt fails (detected by CRC check), a new attempt tries to identify and flip the first bit error occurred during previous attempts.
Further attempts are made until the decoded bits pass CRC check, or a maximum number of attempts is reached.
The bit errors are categorized to two types:
\begin{itemize}
  \item Type-1 as the first error bit, i.e., all its previous bits are correctly decoded.
  \item Type-2 as the subsequent error bits (caused by Type-1 due to error propagation), i.e., there is at least one previous error bit.
\end{itemize}
Clearly, flipping a Type-2 error does not correct a block error, and one should flip the Type-1 error only.

The key to SC flip is to accurately identify Type-1 error bits among the non-frozen bits.
There exist some \emph{heuristic} methods for this purpose.
In \cite{SCFlip}, the bits with smaller amplitude of log likelihood ratio (LLR) are flipped.
However, the amplitude is a sub-optimal metric, since it does not distinguish between Type-1 and Type-2 errors.
Based on this observation, a new metric is adopted in dynamic-SC flip decoding \cite{DynamicFlip} such that the sequential nature of SC decoding is taken into consideration.
Alternatively, the Type-1 errors can be narrowed down to a ``critical set'' to reduce search space \cite{criticalset}.
The critical set can be determined offline in an channel-independent manner to facilitate implementation.
Similarly, the distribution of the first-error is exploited to further reduce search space \cite{error}.
All these papers effectively determine the Type-1 error bits or narrow down the search space, such that the average complexity of SC flip decoding is comparable to SC decoding in high signal-to-noise ratio (SNR) regime.
However, the optimal flipping strategy (which bit to flip) is still an open problem due to the lack of a complete mathematical characterization.

\begin{figure}[]
\centering
\includegraphics[width=3in]{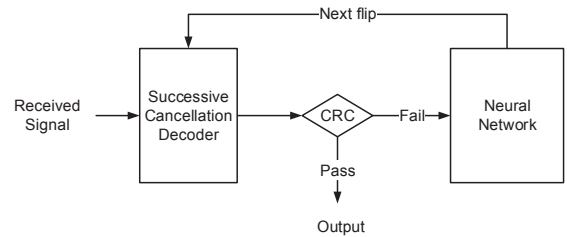}
\caption{The framework of the deep learning aided SC flip algorithm.}
\label{framework}
\end{figure}

When an optimal solution is unavailable, deep learning algorithms are worth trying \cite{DePhysical,nokia1,nokia2,Lyu}.
Recently, deep learning \cite{DeepLearning} has achieved tremendous progress in tasks such as Go\cite{Go}, image classification\cite{image}, machine translation\cite{LSTM}.
Inspired by these, we apply deep learning to identify first error bits during SC flip decoding, as shown in Fig.~\ref{framework}.
Before each SC decoding attempt, a neural network (NN) helps to locate the Type-1 error among the non-frozen bits. The input and output of the network are:
\begin{itemize}
  \item {\bf Input}: a sequence of LLRs (corresponding to both frozen and non-frozen bits) in the previous SC attempt.
  \item {\bf Output}: scores of actions (to flip an error bit), i.e., a vector, each element of which corresponds to the probability of a bit being a Type-1 error. 
\end{itemize}

This is similar to a multi-step maze game, where multiple decisions (Type-1 errors) are made for a right path (successful decoding).
The NN is the game \emph{player} and the LLR sequence is the observed \emph{state} of the maze game.
Since the sequential aspects of SC decoding is already taken into account in the LLR sequence (a latter LLR depends on all the previous LLRs and hard decisions), the NN should have adequate information to locate the Type-1 errors. The remaining questions are:
\begin{enumerate}
  \item What type of NN do we choose?
  \item How to train the NN to identify multiple Type-1 errors?
\end{enumerate}

For the first question, we adopt the long short-term memory (LSTM) \cite{LSTM} network as it is suitable for sequence-based tasks. Given the LLR sequence, the SC flip problem falls into this category. The LSTM network output is further transformed into a vector through a fully-connected layer and a softmax layer. A value in the vector corresponds to the probability of a bit being a Type-1 error.

For the second question, we should keep in mind that even the optimal SC flip algorithm can not guarantee that the first-error bit is always correctly identified.
That means the NN should be trained to not only identify a new flip position, but also undo previous ``wrong'' flips that lead to decoding failures.
From the NN perspective, the previous actions take the player to a \emph{sick} middle state, and the NN should learn to escape from these states.
To achieve this, the \emph{states} and \emph{actions} should be carefully designed to better feed the LSTM network and to support all the required actions (including new flips and undo flips), respectively.
Moreover, we propose a two-stage training method as follows:
\begin{enumerate}
\item Train the network using \emph{supervised learning} to make right actions in the initial states, specifically, to identify the first Type-1 error bit.
\item Continue to train the network using \emph{reinforcement learning} to take right actions in the middle states, specifically, to undo previous ``wrong'' flips and identify subsequent Type-1 error bits.
\end{enumerate}

The contributions are summarized as follows:
\begin{itemize}
\item  A novel deep learning aided SC flip decoding algorithm is proposed.  Compared to the state-of-the-art SC flip algorithms\cite{DynamicFlip}, our algorithm identifies error bit more accurately and achieves better performance.
\item  To train the neural network to make right actions both in initial and middle states, we propose a way to encode the \emph{states} and \emph{actions}. Moreover, a novel two-stage training method that comprises both supervised learning and reinforcement learning is proposed.
\end{itemize}

The rest of the paper is organized as follows. In Section \ref{2}, the LSTM network is briefly introduced. Section \ref{3} discusses network architecture designs. In Section \ref{decoding}, the deep learning aided SC flip decoding algorithm is discussed. Section \ref{4} discusses the two-stage training method and evaluates the performance. Finally, Section \ref{5} concludes the paper.  


\section{LSTM networks}\label{2}
\begin{figure}[]
\centering
\includegraphics[width=3.3in]{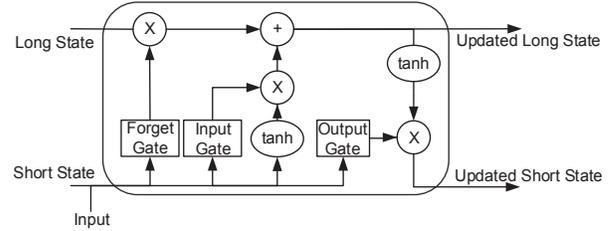}
\caption{A basic LSTM unit.}
\label{Basic_LSTM_cell}
\end{figure}
As shown in Fig.~\ref{Basic_LSTM_cell}, a basic LSTM unit is composed of a cell and three gates, including a forget gate, an input gate and an output gate.
Generally, a long sequence of symbols are partitioned into several subsequences, and fed into the LSTM network one at a time.
Upon receiving each subsequence, the cell stores the state values 
while the gates regulate the information flow into and out of the cell.
With well-trained gates, the cell may store values over arbitrary long time intervals (also called long states).
Specifically, each gate is controlled by a fully-connected multi-layer perceptron (MLP) network and outputs values within $[0,1]$, in which 0 (resp. 1) means the gate is totally closed (resp. open).
The forget gate decides which information to be discarded from the cell and the input gate decides which new input values to be stored into the cell.
Moreover, the output gate decides which information to be output (as short states).
By carefully training the LSTM network, these gates may be properly opened or closed, making it effective in extracting useful features from the sequential structure.
Please refer to \cite{LSTM} for more information about the LSTM network.

\section{Architecture Design}\label{3}
As discussed, SC flip decoding can be viewed as a multi-step maze game with the LSTM as the \emph{player}.
In each step, the \emph{player} chooses an \emph{action} based on the observed \emph{state}.
Thus, the design key is how to encode these \emph{states} and \emph{actions}.

\subsubsection{Encoding of States}
Denote by $\bar{\alpha}$ the LLR sequence (corresponding to both frozen and non-frozen bits) derived in an SC decoder.
We propose to use $|\bar{\alpha}|$, rather than $\bar{\alpha}$, to represent a \emph{state}, due to that the position of the first-error bit mainly depends only on the amplitudes of LLRs.
For example, consider two sequences of $\bar{\alpha}$ with same amplitudes but different signs, their first-error positions are most likely the same.
This is reasonable due to the symmetric channel assumption.
Therefore, $|\bar{\alpha}|$ provides sufficient information.
By removing the ``sign'' information before feeding into the LSTM network, the training complexity can be reduced since less redundant information needs to be processed.

\subsubsection{Encoding of Actions}
One-hot encoding scheme of the action is adopted. \emph{One-hot} means that only one bit in the coded sequence is  $1$ while all the other bits are $0$.
As discussed, a player should be allowed to undo a ``wrong'' flip and go back to the previous state.
We add a bit in the one-hot vector to indicate an ``undo action''.
Therefore, an action is encoded by a binary vector of length $K+1$.
Specifically, the first $K$ bits correspond to the $K$ non-frozen bits, and the last bit corresponds to ``undo'' the previous flipping.

\begin{figure}[]
\centering
\includegraphics[width=2.7in]{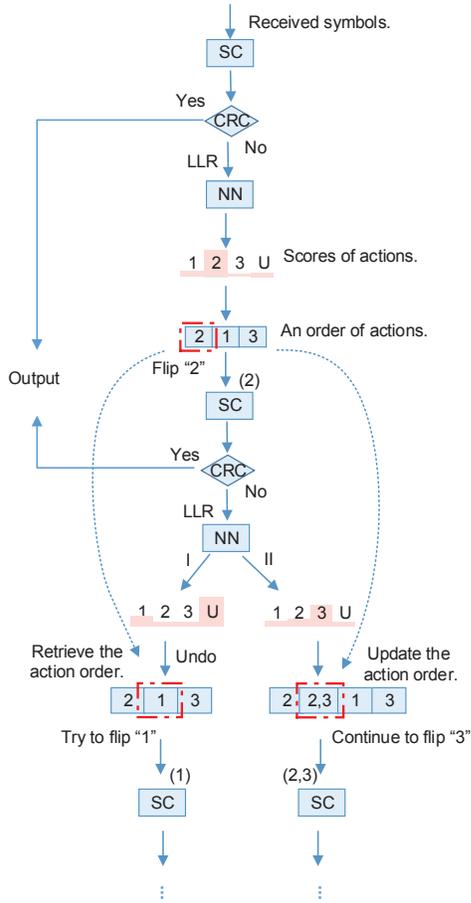}
\caption{The decoding procedure of the deep learning aided SC flip algorithm. At first, the received symbols are decoded with an SC decoder.
If the decoding fails, the LSTM network exploits the LLR sequence to derive an order of actions, i.e., $\{2,1,3\}$.
Then, a new SC attempt tries to flip bit `2'.
If the decoding still fails, the LSTM network is exploited to (I) undo the previous ``wrong'' flip, or (II) identify the next Type-1 error.
Specifically, for Case I, since the last value of the network output, which indicates to ``undo action'', is larger than all the other values, the previous action order is retrieved, based on which, the next SC attempt flips bit `1'.
For Case II, the action order is updated as $\{2,[2,3],1,3\}$ based on the network output, according to which, the next SC attempt flips both bit `2' and bit `3'.
}
\label{flipdecoding}
\end{figure}
\section{Neural Network aided SC flip algorithm}\label{decoding}
The neural network is exploited to aid the SC flip decoding.
As in Fig.~\ref{flipdecoding}, the received symbols are first decoded with an SC decoder.
If it fails, the LSTM network exploits the LLR sequence to score the actions, i.e.,  estimate the probability of each non-frozen bit being a Type-1 error.
Specifically, each LLR sequence is partitioned into a number of subsequences, and then the subsequences are fed into the LSTM network one at a time.
After feeding one LLR sequence to the LSTM network, the final long state is transformed into a vector, to score the actions, through a fully-connected layer and a softmax layer.
Based on the scores, an order of actions are obtained (the details are discussed in Section~\ref{sect:1stErr}).
Following the action order, more SC decoding attempts are conducted as follows. In each attempt, the LSTM network is first exploited based on the LLR sequence (derived in the previous attempt) to either (i) undo a previous ``wrong'' flip, or (ii) identify the next Type-1 error.
Specifically, if the last value of the network output, which corresponds to the ``undo action'', is larger than all the other values, the previous flipping is considered as a ``wrong'' flipping. The next attempt undoes the previous flipping and then tries the next candidate in the action order.
Otherwise, the action order is updated by inserting a ``continue-to-flip action'' according to the action scores. The previous flipping is considered as a ``correct'' flipping and the next attempt continues to flip the next Type-1 error.

\section{Network Training and Performance Evaluation }\label{4}
In this section, we discuss a two-stage training method and evaluate the performance.
As the entire training is conducted off-line, it does not increase the decoding complexity.

Before stepping into details, we describe the loss function that measures the difference between the actual network output and the expected one.
We adopt the \emph{cross-entropy} function, which has been widely used in classification tasks\cite{DeepLearning}, defined as follows:
\begin{equation}\small\label{me_all}
\begin{aligned}
E(\bold{w}) = -\sum_{k}\{{t_k\ln{y_k} + (1-t_k)\ln{(1-y_k)}}\},
\end{aligned}
\end{equation}
where $t_k$ and $y_k$ respectively denote the expected value and the actual value of the $k$-th output.
The smaller the loss function is, the more similar the actual network output is to the expected one, which means the network is better trained.

\subsection{Train LSTM to identify the first-error bit}\label{sect:1stErr}
In the first training stage, a data base is generated, in which the first-error position of each sample has been labeled. 
Specifically, each sample in the data base contains $|\bar{\alpha}|$ (as the input of the neural network) and the one-hot encoding of the first error bit (as the expected output).
To achieve this, we carry out extensive simulations as follows. Random information bits are encoded using polar codes.
Then, the SC algorithm is used to decode the noisy BPSK symbols.  If the decoding result does not verify CRC, there is at least one error bit. In this case, we store the pair of $|\bar{\alpha}|$ and the corresponding one-hot representation of the first-error bit as a sample in the data base.

The data base is then used to train the neural network to minimize the loss function.
To properly train the network, we divide the data base into two sets: a training set and a validation set.
The training set is used to train the network, while the validation set is used to avoid overfitting.
Note that the training quality is largely influenced by certain hyperparameters, such as size of training set, mini-batch size and so on.
In this paper, we do not focus on the optimization of these hyperparameters and just choose a set of hyperparameters that works.

\subsubsection{Performance Evaluation}
We evaluate the performances for polar codes of length $N=64$ or $128$.
In the simulation, the coding rate is $\frac{1}{2}$ and the CRC length is $8$.
Both the training and testing SNR are set to $1$dB. 
The hyperparameters are listed in Table~\ref{tab1}.
\begin{table}[H]
\caption{Hyperparameters Set I}
\begin{center}
\begin{tabular}{c|c}
\hline
LSTM network & 3 layers with hidden-size of 256\\
\hline
Size of Training set & $1.6\times10^6$ \\
\hline
Size of Validation set &$5\times10^4$\\
\hline
Mini-batch size & $10^3$\\
\hline
Subsequence size & $4$\\
\hline
Number of epochs & 30\\
\hline
Dropout & 0.05\\
\hline
Optimizer & Adam\\
\hline
\end{tabular}
\label{tab1}
\end{center}
\end{table}

\begin{figure}[]
\centering
\includegraphics[width=3.2in]{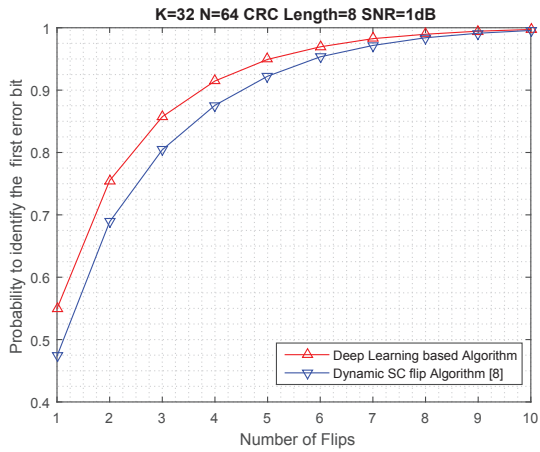}
\caption{For the Polar code of $N=64$, the proposed deep learning aided algorithm is more accurate to identify the first-error bit.}
\label{F1R}
\end{figure}

\begin{figure}[]
\centering
\includegraphics[width=3.2in]{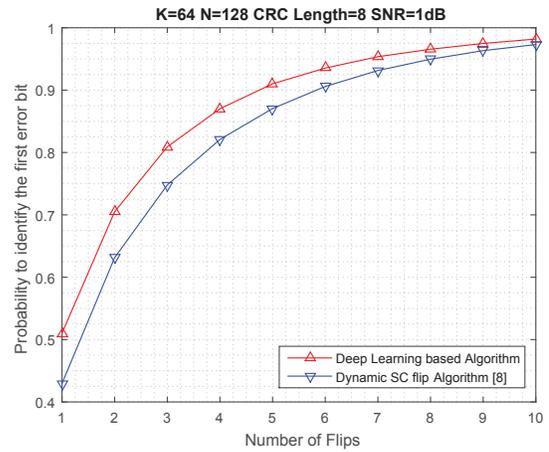}
\caption{For the Polar code of $N=128$, the proposed deep learning aided algorithm is more accurate to identify the first-error bit.}
\label{F2R}
\end{figure}

The simulation results are presented in Fig.~\ref{F1R} and Fig.~\ref{F2R}.
As a benchmark, the performance of the state-of-the-art SC flip algorithm, i.e., dynamic SC flip algorithm \cite{DynamicFlip}, is presented.
To ensure a fair comparison, the parameter $\alpha$ in the dynamic SC flip algorithm is optimized with the method in \cite{DynamicFlip}.
The probability to successfully identify the Type-1 error within the same number of flips is compared.
According to the simulation results, the proposed deep learning aided algorithm achieves higher successful rates than the benchmark algorithm.

\begin{table}[H]
\caption{The rate to successfully identify the first error bit in different attempts.}
\begin{center}
\begin{tabular}{|c|c|c|c|c|}
\hline
 & First & Second& Third & Fourth\\
\hline
N=64 & 0.547 & 0.207 & 0.102 & 0.055\\
\hline
N=128 & 0.510 & 0.192 & 0.098 & 0.053 \\
\hline
\end{tabular}
\label{tab3}
\end{center}
\end{table}
Moerover, the rates to successfully identify the Type-1 error in different attempts are provided in Table~\ref{tab3}.
As shown, the LSTM network not only identifies the most likely first-error bit position, but also outputs other candidate bit flip positions by descending probability order.
This means that, if a bit flip action fails, the LSTM can provide an order of actions for bit flipping in the subsequent attempts.
This is crucial for multi-bit successive flipping, to be discussed shortly.


\subsection{Train LSTM to undo previous ¡°wrong¡± flips}
Consider the case in which the SC attempt still fails after a bit flip.
The failure is associated with two reasons: 1) the first-error bit is wrongly identified; 2) the first-error bit has been correctly flipped, but there are additional error bit(s) in the following.
The first case cannot be avoided completely, since, as discussed, the neural network may not always find the Type-1 error bit. A ``wrong'' flip takes the player to a \emph{sick} middle state, in which the network has not been trained during the first training stage. Therefore, the neural network should be further trained to escape from these states, i.e., undo the ``wrong'' flips.
Since the middle states depend on the previous actions (decided by the neural network), it is difficult to train the network in these states using supervised learning.

\begin{figure}[]
\centering
\includegraphics[width=2.6in]{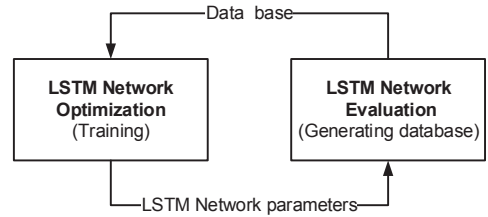}
\caption{LSTM network is trained iteratively using \emph{reinforcement learning} to flip multiple error bits successively. In each iteration, a data base that comprises three types of states, i.e., initial states, general middle states and sick middle states, is generated. Based on the new data base, the LSTM network is trained. Specifically, a well-trained LSTM network to identify the first Type-1 error (as discussed in Section~\ref{sect:1stErr}) is taken as the starting point, so as to accelerate the training process.
}
\label{RFL_framewrok}
\end{figure}

To address the issue, in the second training stage, the LSTM network is further trained iteratively, as shown in Fig.~\ref{RFL_framewrok}. In each iteration, a new data base is generated based on the LSTM network. Then, the LSTM network is trained using \emph{reinforcement learning} based on the new data base.

To generate the data base, we carry out extensive simulations as follows.
Polar codes are decoded with the proposed deep learning aided SC flip algorithm.
Upon encountering \emph{sick} middle states, a one-hot vector with only the last bit as $1$ is set as a label.
For other states, the labels are set as in Section~\ref{sect:1stErr}.
In this way, we obtain a data base that comprises three types of states, i.e., initial states, general middle states and \emph{sick} middle states.

Note that the samples in the data base may be grouped into sequences, in which, the latter ones depend on the previous ones.
As such, when applying it to the network training, the sum loss for a sequence can be considered as long-term loss. Therefore, the training process is reinforcement learning.

\subsection{Performance Evaluation}
\begin{table}[H]
\caption{Hyperparameters set II}
\begin{center}
\begin{tabular}{c|c}
\hline
Size of Training set & $1.6*10^6$ \\
\hline
Size of Validation set &$5*10^4$\\
\hline
Mini-batch size & $10^3$\\
\hline
Sub-sequence size & $4$\\
\hline
Number of epochs & 20\\
\hline
Number of iterations & 60\\
\hline
Dropout & 0.05\\
\hline
Optimizer & Adam\\
\hline
\end{tabular}
\label{tab2}
\end{center}
\end{table}

We evaluate the performances of the proposed algorithm for polar code of length $N=64$. The coding rate is $\frac{1}{2}$ and the CRC length is $8$.
Both the training SNR and testing SNR are set to 1dB and the hyperparameters are listed in Table~\ref{tab2}.
\begin{figure}[]
\centering
\includegraphics[width=3.2in]{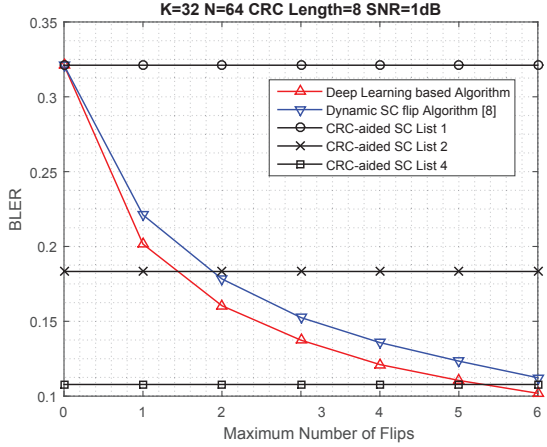}
\caption{The proposed algorithm achieves better BLER performance than the dynamic flip algorithm. With at most $5$ attempts, the proposed algorithm approaches the performance of CA-SC-List-$4$ algorithm. As the number of attempts increases, the BLER decreases slower, which is due to that the extra decoding attempts are performed with descending success probability. }
\label{BLER}
\end{figure}
The state-of-the-art SC flip algorithm, i.e., dynamic SC flip algorithm \cite{DynamicFlip}, is taken as the benchmark.
According to the simulation results in Fig.~\ref{BLER}, with the same number of flips, the proposed algorithm achieves better BLER performance than the dynamic SC flip algorithm.
Moreover, the proposed algorithm is also compared to the CA-SC-List algorithm, which shows that, with at most $5$ attempts, our algorithm approaches the performance of CA-SC-List-$4$ algorithm.

\section{Conclusion and Future Work}\label{5}
In this paper, we apply the LSTM networks to identify error bits under SC decoding, based on which, we propose a framework of deep learning aided SC flip algorithm for polar codes. 
The framework offers a new method to exploit deep learning in decoding algorithms, and may be applied to polar codes of moderate lengths.
To train the neural network to successively flip multiple error bits, we propose a two-stage training method that comprises both supervised learning and reinforcement learning.  Evaluation results show that the proposed approach identifies error bits more accurately and achieves better block error rate performance than the state-of-the-art SC flip algorithm.

To further improve the algorithm, several open problems and possible directions are listed as follows:
\begin{itemize}
\item The same training and testing SNR is evaluated in this paper. The performance for the case with SNR mismatch needs further analysis and evaluation.
\item The LSTM network has been trained without human knowledges in this paper.  Human knowledges, e.g., critical set\cite{criticalset}, may help improving the performance further.
\end{itemize}

\bibliographystyle{IEEEtran}

\end{document}